\documentclass{PoS}

\title{Multifrequency Behaviour of Galactic and Extragalactic Novae}

\ShortTitle{Multifrequency Behaviour of Galactic and Extragalactic Novae}

\author{\speaker{Rosa Poggiani}\\
       Universit\`a di Pisa and Istituto Nazionale di Fisica Nucleare, Sezione di Pisa\\
       E-mail: \email{rosa.poggiani@df.unipi.it}}

\abstract{The outbursts of novae are among the strongest explosions in the Universe. The eruptions involve physical processes that span the whole electromagnetic spectrum, demanding multifrequency observations. The photometric and spectroscopic observations in the optical domain are combined with radio, infrared, X-ray and gamma ray observations, allowing to discover new phenomena, like the GeV gamma ray emission.}

\FullConference{XII Multifrequency Behaviour of High Energy Cosmic Sources Workshop\\
         12-17 June\\
         Palermo, Italy}

\begin{document}

\section{Introduction}

Classical novae are cataclysmic variable systems, composed of a white dwarf that is accreting material from a secondary star \cite{PayneGaposchkin1957}, \cite{CassatellaViotti1990}, \cite{BodeEvans2008}. The eruptions of novae, marked by the rapid increase in optical luminosity, are the result of a ThermoNuclear Runaway (TNR) occurring on the surface of the white dwarf \cite{CassatellaViotti1990}, \cite{BodeEvans2008}, \cite{Starrfield2016}. The systematic investigation of classical novae is relatively recent, dating back to the end of Nineteenth Century \cite{Duerbeck2009}. A large part of historical observations is in the optical domain, that presently can rely on high cadence photometric and spectroscopic observations. Novae are suitable objects also for multifrequency investigation with large ground based and satellite based instruments, since they emit over the whole electromagnetic spectrum: a multifrequency picture of the nova process can be built combining the optical observations with high energy and radio ones \cite{Giovannelli2015}. Novae are of general interest since they are a suitable arena to study the accretion processes and can also be used as standard candles. A general introduction to the topics related to cataclysmic variables and novae can be found in \cite{Giovannelli2015}.

In this paper, I review some aspects of the phenomenology of Galactic and extragalactic novae in the context of multifrequency observations. Firstly, the basic data about Galactic and extragalactic novae will be presented to provide a starting point for discussion. Then the optical photometric and spectroscopic observations, the pillars of nova classification, will be presented. The observations in the different regions of the electromagnetic spectrum are discussed in the following sections. The possible gravitational emission of novae is discussed in the last section.

\section{Novae: the basic facts}
 
Novae are usually classified into classical novae and recurrent novae, according to the number of observed eruptions, one or more. The predicted recurrence times range from tens to hundred years for recurrent novae and is above tens thousand years for classical novae.
The interval between outbursts is mostly determined by the mass of the white dwarf and by the accretion rate \cite{Yaron2005}.  Recurrent novae have an high rate of mass transfer that explains the short interval between outbursts \cite{Orio2015}. Recurrent novae are possible progenitors for type Ia supernovae \cite{Surina2011}, \cite{Shafter2015}. 
 
The typical signature of a nova is the rapid rise to the maximum optical brightness, followed by a decline. Novae are classified using observations in the optical domain (see next sections), that suggest two distinct spectral classes, the Fe II novae and the He/N novae, according to the most intense lines besides the Balmer ones \cite{Williams1991}, \cite{Williams1992}, \cite{Williams1994}. He/N novae show high ejection velocities and fast declines. On the other hand, Fe II novae exhibit low to high ejection velocities and either fast or slow declines. As discussed in Sec. \ref{spectroscopic}, all novae go through both He/N and Fe II phases, but with different durations, according to \cite{Williams2012}.

The early decline observed in the optical domain is accompanied by a rise of the ultraviolet component. The presence of a dip produced by dust formation causes an increase of the infrared emission. During the later stages, radio and X-ray emission appear, with the former lasting for longer times. A model addressing multifrequency evolution has been proposed by \cite{HachisuKato2006}, \cite{KatoHachisu1994}.
Most of the envelope is expelled by winds. After achieving the maximum expansion, the photosphere begins to shrink, with ejecta still undergoing expansion. Ejecta that are optically thin emit free-free radiation, while the photosphere is dominated by blackbody-like emission. 

Galactic novae have been cataloged by \cite{Duerbeck1987}, \cite{Downes2005}, \cite{RitterKolb2003}, \cite{RitterKolb}; a live catalog of the most recent novae is maintained by \cite{Mukai}. To date, more than four hundreds Galactic novae are known.

The rate of Galactic novae is a key parameter for the understanding of the nova events and for the chemical evolution of the Galaxy, since novae can contribute with $^7$Li, $^{15}$N, $^{22}$Na, $^{26}$Al. An investigation of the galactic nova rate has been presented by \cite{Shafter2017}. Novae are preferentially concentrated close to the Galactic plane, in particular close to the Galactic center, with a bias for observation of brighter novae in the Northern hemisphere. The estimated galactic nova rate is 50$^{+31}_{-23}$ yr$^{-1}$. For comparison, other estimations of the rate are reported in Table \ref{tab:galrate}.

\begin{table}[h]
\centering
\tabcolsep=0.11cm
\scalebox{0.9}{
\begin{tabular}{lr} \hline
Authors		& Rate (yr$^{-1}$)	\\ \hline
Allen 1954  \cite{Allen1954}		&  100 \\
Sharov 1972 \cite{Sharov1972}   &  260 \\ 
Liller and Mayer 1987 \cite{LillerMayer1987}  & 73$\pm$24 \\
Della Valle 1988 \cite{DellaValle1988}        & 15$\pm$5  \\
Ciardullo et al. 1990 \cite{Ciardullo1990}    & 11 to 46  \\ 
van den Bergh 1991 \cite{vandenBergh1991}     & 16 \\ 
Della Valle and Livio 1994 \cite{DellaValleLivio1994}  & 20 \\ 
Hatano et al. 1997 \cite{Hatano1997}          & 41$\pm$20  \\ 
Shafter 1997 \cite{Shafter1997}               & 35$\pm$11  \\ 
Shafter 2002 \cite{Shafter2002}               & 36$\pm$13  \\ 
Matteucci et al. 2003 \cite{Matteucci2003}    & 25 \\ 
Mroz 2015 et al. \cite{Mroz2015}              & 13.8$\pm$2.6 \\
Shafter 2017 \cite{Shafter2017}               & 50$^{+31}_{-23}$ \\
\hline
\end{tabular}
}
\caption{Compilation of estimated galactic nova rates reported in literature}
\label{tab:galrate}
\end{table}

\begin{figure}[h]
\centering
\includegraphics[width=.6\textwidth]{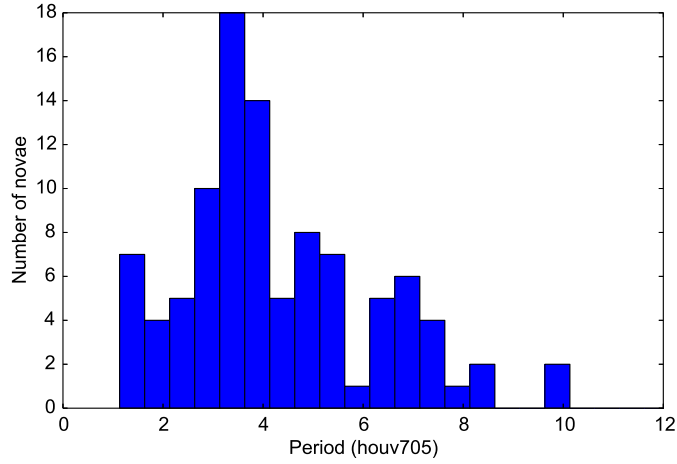}
\caption{Distribution of the orbital periods of Galactic novae, data from \cite{RitterKolb}}
\label{fig:norb}
\end{figure}

The distribution of orbital periods built using the data by \cite{RitterKolb2003}, \cite{RitterKolb} is reported in Fig. \ref{fig:norb}, where novae with evolved secondary stars and periods exceeding 10 hours have been excluded. The orbital period has been measured only for a small fraction of the observed novae. The distribution shows a dominance of systems above the period gap of cataclysmic variables, with a peak in the region between 3 and 4 hours, as discussed by \cite{Tappert2015}. Novae are mostly systems with a high mass accretion rate onto the white dwarf \cite{Tappert2015}.

The relatively high luminosity of novae makes them observable in close galaxies, such as M31, M33, LMC, avoiding the problem of variable interstellar extinction in the Milky Way. A review of extragalactic novae has been presented by \cite{Shafter2014}. One subject of investigation of extragalactic novae is the connection with the stellar populations in their galaxies and its impact on the rate and evolution. The rate of novae in galaxies spanning different types has been tackled by \cite{Yungelson1997} using population synthesis techniques. A parameter governing the nova rate in a galaxy is the evolution of its star formation rate (SFR). The nova rate is usually re-scaled to unit mass using the K-band luminosity for comparison. The re-scaled rate is higher for less massive galaxies and lower for elliptical galaxies, where star formation is concentrated in an initial burst.
An investigation of theoretical nova rates in galaxies belonging to different Hubble types has been performed by \cite{Matteucci2003}. 
Other possible differences in nova properties are related to the site of nova event, in disk or in bulge: bulge novae belong to an older stellar population that is related to less massive white dwarfs.

The main contributions to the statistics of extragalactic novae is given by the close galaxies M31, M33, LMC, that will be discussed in the following. The catalogs of extragalactic novae \cite{Pietsch2010}, updated at June 2017, contain 1071 novae in M31 \footnote{http://www.mpe.mpg.de/$\sim$m31novae/opt/m31/index.php}, 50 in M33 \footnote{http://www.mpe.mpg.de/$\sim$m31novae/opt/m33/index.php}, 50 in LMC \footnote{http://www.mpe.mpg.de/$\sim$m31novae/opt/lmc/index.php}.

The nova rate in M31 has been investigated for a long time, beginning with the first photographic surveys. A compilation of rates is presented in Table \ref{tab:m31rate}. Due to saturation of plates, the photographic surveys \cite{Hubble1929}, \cite{Arp1956}, \cite{Rosino1964}, \cite{Rosino1973}, \cite{Rosino1989} were missing novae occurring close to the galaxy nucleus \cite{Capaccioli1989}. Novae can in fact appear in any part of M31 \cite{Ciardullo1987}, most of them being concentrated in the halo and in the bulge \cite{Capaccioli1989}.

\begin{table}[h]
\centering
\tabcolsep=0.11cm
\scalebox{0.9}{
\begin{tabular}{lr} \hline
Authors		& Rate (yr$^{-1}$)	\\ \hline
Hubble 1929 \cite{Hubble1929}	& 30 \\ 
Arp 1956 \cite{Arp1956}		& 26 \\	
Capaccioli et al. 1989 \cite{Capaccioli1989}	& 29$\pm$4 \\
Shafter and Irby 2001 \cite{ShafterIrby2001}  & 37$^{+12}_{-8}$ \\
Darnley et al. 2006 \cite{Darnley2006}      & 65$^{+16}_{-15}$ \\
\hline
\end{tabular}
}
\caption{Compilation of estimated nova rates in M31 reported in literature}
\label{tab:m31rate}
\end{table}

The rate of novae in M33 is predicted to be larger than in M31, since M33 is an early galaxy \cite{DellaValle1994}.
The estimated rate ranges from less than 0.4 yr$^{-1}$ \cite{Sharov1993} to  4.7$\pm$1.5 yr$^{-1}$ \cite{DellaValle1994}, including the recent estimated  value 2.5$^{+1.0}_{-0.7}$ yr$^{-1}$  \cite{Williams2004}.

The rate in LMC is of the order of a few novae per year:
2 to 3 novae per year \cite{Graham1979}, 2$\pm$1 yr$^{-1}$ \cite{Capaccioli1990}, 2.5$\pm$0.5 yr$^{-1}$ \cite{DellaValle2002},
2.4$\pm$0.8 yr$^{-1}$ \cite{Mroz2016a}. The K-band luminosity specific nova rate is higher than in other galaxies, probably due the peculiar star formation history that showed a second burst.

\section{Photometric evolution} 

\begin{figure}
\centering
\includegraphics[width=.6\textwidth]{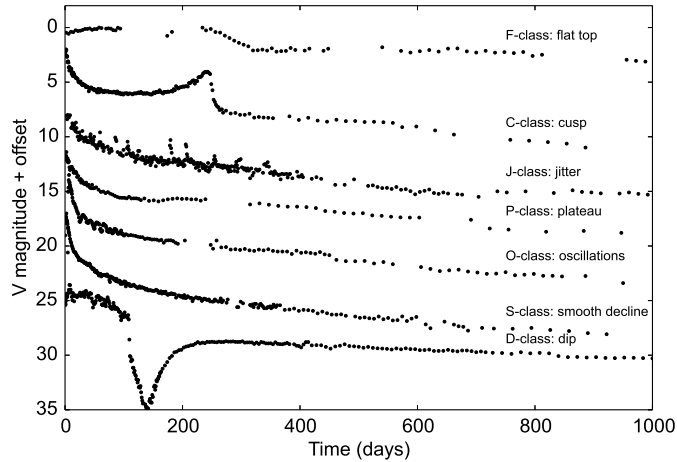}
\caption{The classification of the novae light curves, after \cite{Strope2010}; data from \cite{Strope2010}}
\label{fig:curves}
\end{figure}

The light curve of novae during the decline are not always smooth, but peculiar features such oscillations, dips, cusps, flares can appear. The photometric behavior of novae has been systematized by \cite{Strope2010}, who built a catalog of 93 light curves and devised a classification system. There are seven prototype classes, marked by the features exhibited during the decline: the F-class novae with flat top (2\% of all novae); the C-class novae have a cusp (1\%); the J-class novae show jitters (16\%); the P-class novae show a plateau (21\%); the O-class novae show oscillations (4\%); the S-class systems show a featureless decline (38\%); the D-class novae show dips (18\%). A smooth decline can be expected only for less than one half of novae. A synoptic view of the decline curves, modeled after \cite{Strope2010}, is reported in Fig. \ref{fig:curves}. Also extragalactic novae can show the non standard photometric and/or spectroscopic behaviors sometimes observed in Galactic novae.

The model by \cite{HachisuKato2006}, \cite{KatoHachisu1994} predicts a universal decline law for novae that do not show peculiar features (cups, dips, oscillations..), for a range of compositions and masses of the white dwarf, after rescaling by a suitable time factor. The universal light curve behaves as t$^{-1.75}$ in the time range corresponding to the drop from to 2 to 6 magnitudes below the optical maximum and as t$^{-3.5}$ in the time region corresponding to the drop from 6 to 10 magnitudes below maximum, where t is the time elapsed from the outburst in days. 

Luminous novae are often discovered and monitored by citizen astronomers. Photometric data for most novae are available in the archives of AAVSO \footnote{https://www.aavso.org/} and VSNET \footnote{http://ooruri.kusastro.kyoto-u.ac.jp/mailman/listinfo/}. The last decade has witnessed the operation of all-sky surveys aimed to monitor the variability of known and new sources with high temporal cadence. The surveys discover and monitor a large number of transient sources, whose variability can be searched in archival data. The high cadence of observations has allowed to investigate the orbital period of novae and to monitor the often missed pre-maximum stage. The ground based Optical Gravitational Lensing Experiment (OGLE) has monitored several novae in the Galaxy \cite{Mroz2015}. The survey of Galactic novae by Mroz et al. \cite{Mroz2015} has observed the eruptions of 39 novae (one third of them discovered by the survey), detected 80 post-nova candidate systems and estimated the orbital periods of 18 novae. The space based Solar Mass Ejection Imager (SMEI) instrument performs photometric observations with a cadence of about 100 minutes. The instrument has measured the light curves of 13 galactic novae \cite{Hounsell2010}, \cite{Hounsell2016} and has investigated the poorly known pre-maximum stage for some of them.  Some novae observed by SMEI were discovered by ground based observatories several weeks after the maximum, indicating a possible missing of novae eruptions, even for bright novae, as discussed by \cite{Shafter2017}. 

While the maximum and the decline stages of novae are well covered, the pre-eruption and the long term behaviour of novae are still poorly understood: photometric and spectroscopic observations are at best sporadic before the maximum and decrease in cadence when novae enter the nebular stage. The long term evolution of novae has been discussed by \cite{Patterson2014}.  
The first investigation about the pre-eruption photometry of novae has been performed by \cite{Robinson1975}. Among the 11 studied novae, five of them exhibited a slow brightness rise in the years preceding the outburst; most systems showed the same quiescent magnitude before and after the eruption event. A recent investigation by \cite{Collazzi2009} included 22 novae.  The pre-eruption brightness increases where less frequent than expectations, being confirmed only for two systems. Most novae did not exhibit variations of the quiescent magnitude before and after the eruption. It has been suggested that novae can undergo hibernation between successive eruptions \cite{Shara1986}, due to the lack of observations supporting an high mass transfer rate  maintained more than a century after the outburst.
Mass loss dominates during the nova outburst, separating the secondary star from its Roche lobe and switching off the mass transfer after irradiation. During the hibernation phase, the combination of magnetic braking and gravitational wave emission decreases the distance of the nova components, finally re-establishing the contact with the Roche lobe. The model predicts that old novae show a relatively high brightness lasting for decades, due to the irradiation by the white dwarf. When the irradiation fades, the accretion rate slows down. The hibernation model predicts a cyclic evolution between the two stages stages with high and low mass transfer. When the mass transfer rate slows down, the nova enters a stage where dwarf nova outbursts appear, while the total brightness decreases. The hibernation stage can be as long as hundreds thousands or million years. The model predicts that the majority of novae spend more than 90\% of their lives as detached and close binaries. The hibernation hypothesis is supported by the measurement of the decline rate of old novae  \cite{Duerbeck1992} and by the observation of shells around the dwarf novae Z Cam \cite{Shara2007}, AT Cnc \cite{Shara2012}. The presence of shells demonstrates that some dwarf novae have undergone nova eruptions in the past, and some novae can become dwarf novae after the thermonuclear outburst, as predicted by the hibernation model \cite{Shara1986}.
The observations of nova V1213 Cen performed by \cite{Mroz2016b} during the pre-maximum and the outburst stages have showed the awakening of the object from the hibernation phase. After the outburst that occurred in 2009, V1213 Cen showed a slow fading by about one half a magnitude per year until 2016. The pre-eruption data showed the presence of some dwarf nova outbursts, with a duration of one to two weeks, a spacing of two or three weeks and an amplitude of about three magnitudes,  as in U Gem systems. The brightness in quiescence is compatible with the brightness of dwarf novae. The nova eruption of V1213 Cen occurred less than a week after the onset of a dwarf nova outburst.

The study of novae in quiescence, necessary to understand the conditions that produce the outburst and the possible variations after the eruption, has received a great deal of attention. The systems marked as old novae in the catalogs of cataclysmic variables \cite{Downes2005} do not necessarily have an identified optical counterpart with spectra. A large effort mining in the archive of astronomical photographic plates for novae and transient or flaring events in general, is ongoing. The plate archives of world observatories contain records of astronomical data that can extend for more than one century, but not all plates have been digitized so far. A search for old novae and flaring cataclysmic variables, blazars and X-ray binaries in the plates of several observatories is being performed by \cite{Hudec2015}. A search in the plates of DASCH has been presented by  \cite{Pagnotta2017}. The observational search for old novae is based on multicolour photometry and spectroscopy \cite{Tappert2012}, \cite{Pagnotta2017}. Old novae are selected by their position in the color-color diagram, since the combined effect of the white dwarf, red dwarf and accretion disk shifts it away from the main sequence curve \cite{Tappert2012}. The spectroscopic observations are used for the classification of the object as a nova or a generic cataclysmic variable.

Before the beginning of recent surveys, sparse coverage existed for extragalactic novae. The Magellanic survey \cite{Mroz2016a} has monitored 20 eruptions of novae in LMC and SMC, one half of them original discoveries. The Palomar Transient Factory (PTF) has monitored 29 novae in M31 \cite{Cao2012}. Eight novae of the sample have also been monitored in the near-ultraviolet using the Galaxy Evolution Explorer (GALEX). Two novae, M31 2009-10b and M31 2010-11a, showed  UV emission peaking some days before the optical maximum, possibly due to an aspherical outburst.

\section{Spectroscopic evolution}
\label{spectroscopic}

\begin{figure}
\centering
\includegraphics[width=.3\textwidth]{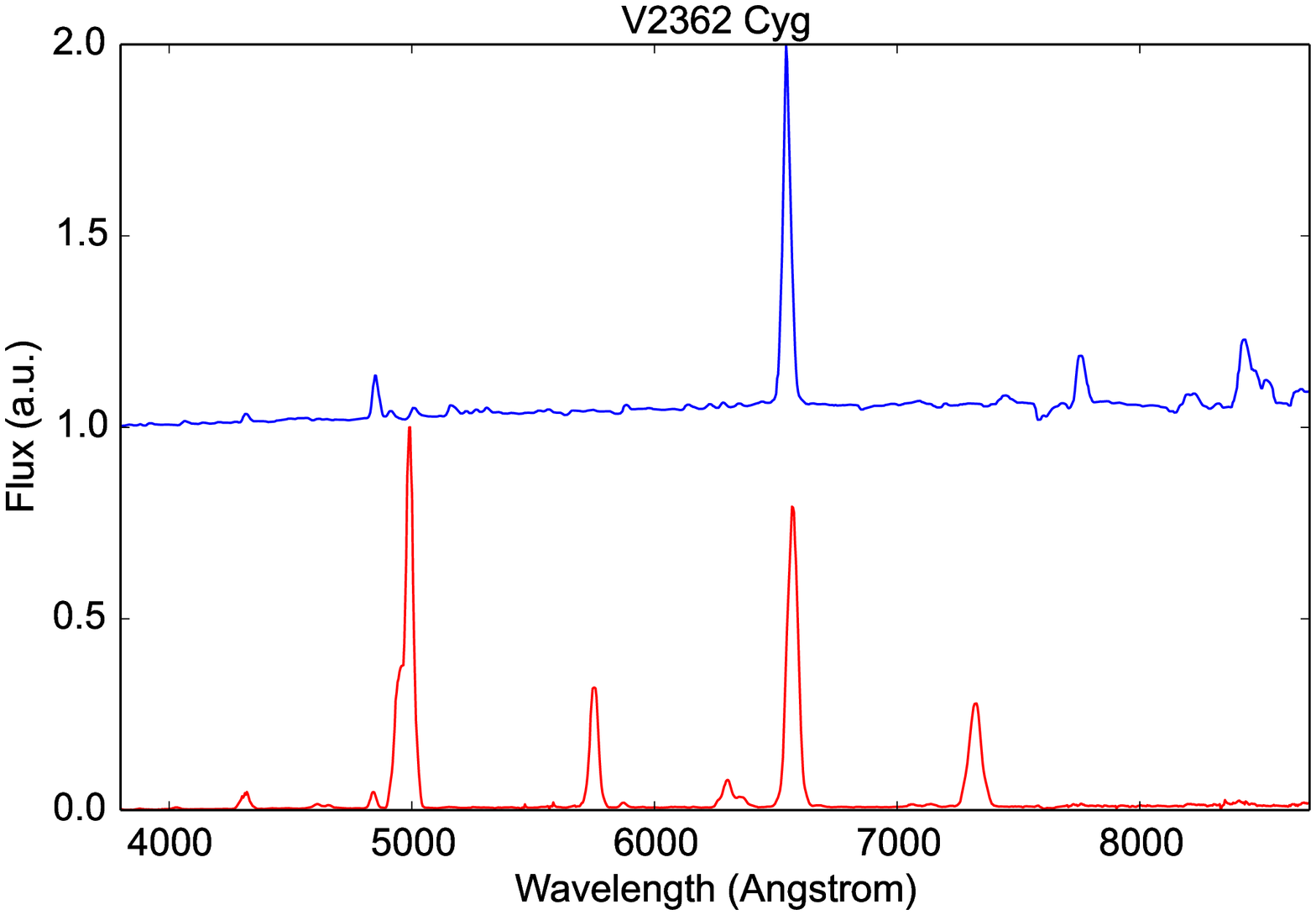}
\includegraphics[width=.3\textwidth]{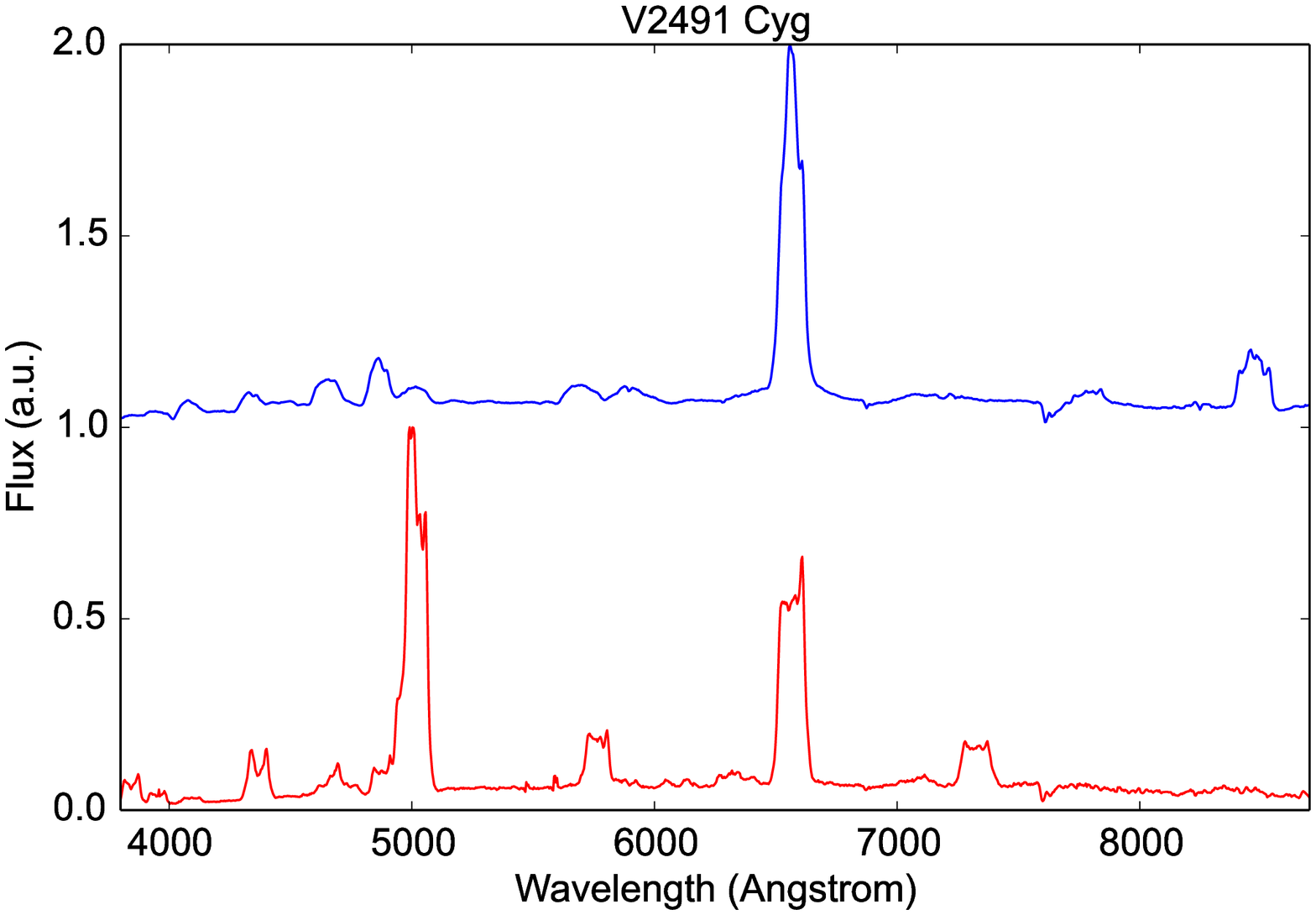}
\includegraphics[width=.3\textwidth]{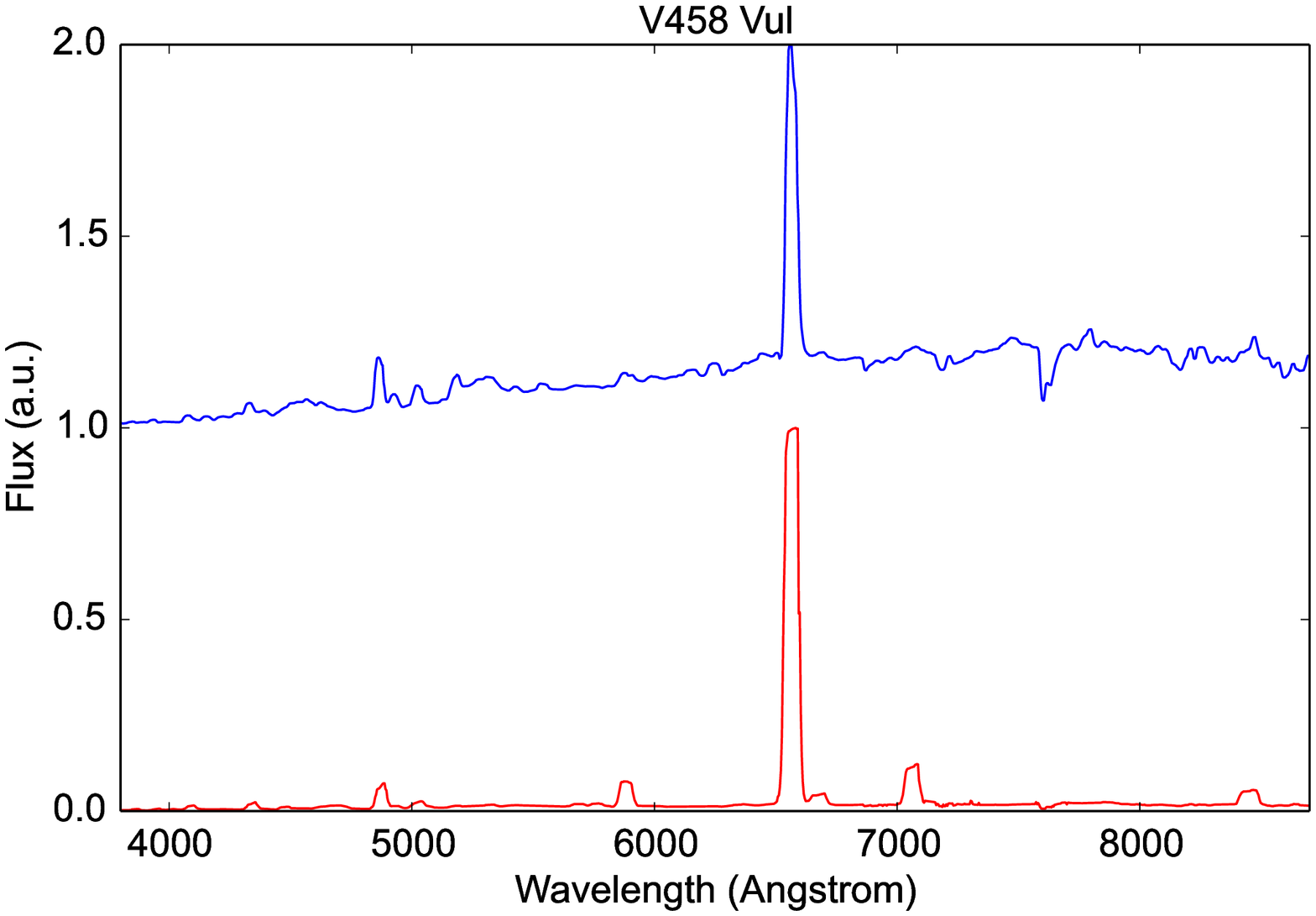}
\caption{Spectra of novae of different spectral classes: the Fe II nova V2362 Cyg \cite{Poggiani2009a}, the He/N nova V2491 Cyg (Poggiani, in preparation), the hybrid nova V458 Vul \cite{Poggiani2008b}}
\label{fig:spectra}
\end{figure}

The spectra of novae can be classified into two main classes, Fe II and He/N novae \cite{Williams1991}, \cite{Williams1992}. Fe II novae show a slow to moderately fast decline and Fe II lines as the strongest non Balmer lines in early spectra, replaced by auroral and forbidden lines in the nebular stage. The physical emission mechanism is wind ejection. Fe II novae account for about 80\% of Galactic novae. He/N novae show a fast decline and broad He or N lines as the strongest non Balmer lines in the early spectra, replaced by coronal lines in the nebular stage. The physical emission mechanism is shell  ejection. 
The spectral classes are related to the two distinct nova populations in the Milky Way: fast novae are closer to the galactic plane than slow novae \cite{DellaValle1992}. He/N and Fe II \cite{Williams1991}, \cite{Williams1992} differ also in the maximum brightness, the former are not only faster but also more luminous than the latter \cite{DellaValleLivio1998}. He/N are more concentrated around the Galactic plane, while Fe II novae can be as far as 1 kpc from the plane \cite{DellaValleLivio1998}. A handful of novae are classified in the hybrid class, since their spectra show an evolution from the Fe II class to He/N class.
The early and nebular spectra of V2362 Cyg (Fe II), V2491 Cyg (He/N) and V458 Vul (hybrid) are reported in Fig. \ref{fig:spectra}.

The recent eruptions of V5558 Sgr \cite{Tanaka2011} and T Pyx (see e.g. \cite{ImamuraTanabe2012}, \cite{Ederoclite2014}) have shown that some novae could also switch from an He/N spectrum to a Fe II spectrum. The suggested mechanism is the combination of weak winds and shell ejection. A global framework has been proposed by \cite{Williams2012}. The physical conditions leading to He/N and Fe II spectra suggest an origin in the white dwarf ejecta or in a circumbinary gas envelope whose origin is the secondary star. Novae where the two classes of spectra are observed in sequence can be modeled with a variable contribution of the two mechanisms during the decline stage. Most novae are expected to show both kinds of spectra during their early evolution.

The complexity of the nova phenomenon has triggered the spectroscopic investigation of large samples of novae. The SMARTS Consortium \cite{Walter2012}, \cite{Walter2014} has prepared a spectral atlas of Southern novae, continuously updated, as the successor of the Tololo nova atlas \cite{Williams1994}, \cite{Williams2003}. In 2017 the atlas contained the optical/IR photometric and spectroscopic observations of some tens novae, secured using the ANDICAM dual channel imager and the RC spectrograph at the SMARTS 1.5 m telescope \cite{Walter2012}. The atlas has been used for a number of synoptic investigations \cite{Walter2014}. Novae can form dust, that is not necessarily accompanied by a dip in the light curve. The K$_s$ band flux shows an increase before the onset of the dip, also for novae where dust formation is not marked by a brightness drop. The combination of the SMARTS monitoring with Swift observation has shown that He II 4686 in fast He/N novae exhibits a temporary increase just before the starting of the super-soft phase (see below), fading at the turn-on.

Since 2005 I am monitoring Northern novae at the Cassini 1.5m telescope, using the BFOSC Imager/Spectrograph \cite{Poggiani2012}. Optical spectra are secured during the  decline with the densest possible sampling. The sample includes more than twenty Galactic novae and some extragalactic novae: V1663 Aql \cite{Poggiani2006}, V1722 Aql, V809 Cep, V962 Cep, V2362 Cyg \cite{Poggiani2009a}, V2467 Cyg \cite{Poggiani2009b}, V2468 Cyg, V2491 Cyg, V2659 Cyg, V407 Cyg, V339 Del, KT Eri, V959 Mon, V2615 Oph, V2670 Oph \cite{Poggiani2009c}, V2944  Oph, V496 Sct, V556 Ser, V5558 Sgr \cite{Poggiani2008a}, \cite{Poggiani2010a}, V5584 Sgr \cite{Poggiani2011}, \cite{Poggiani2015a}, V458 Vul \cite{Poggiani2008b}, V459 Vul \cite{Poggiani2010b}, the extragalactic novae M31 2009-10b, M31 2010-07a, M31 2011-07 and M33 2010-07a \cite{Poggiani2015}. Some novae showing a peculiar photometric and spectroscopic behavior have been investigated. V2362 Cyg showed a secondary brightening during the decline, corresponding to a secondary mass ejection \cite{Poggiani2009a}. V2467 Cyg exhibited oscillations during the decline, with an early appearance of nebular lines \cite{Poggiani2009b}. The very slow nova V5558 Sgr hosts a white dwarf with a mass at the lower limit to trigger a nova outburst \cite{Poggiani2008a}, \cite{Poggiani2010a}. V458 Vul is an hybrid nova that switched from the Fe II to the He/N class after the outburst \cite{Poggiani2008b}.  M31 2009-10b is one of the most luminous novae observed in M31 \cite{Poggiani2015}, while M33 2010-07a is the first nova in M33 exhibiting a secondary mass ejection \cite{Poggiani2015}.

Spectroscopic observations of novae are continuing to produce a wealth of information about the nova phenomenon. An example is given by the early spectra of nova V1369 Cen \cite{Izzo2015}. The presence of lithium in the spectra, combined with the galactic nova rate, can explain the observed over abundance of lithium in young star populations.

The close galaxies M31, M33, have been the subject of targeted spectroscopic investigations. The possibility of two nova populations in M31, in bulge and in disk, has been suggested by \cite{Darnley2006}.
A multi-year spectroscopic investigation of 46 novae in M31 has been performed by \cite{TomaneyShafter1992}, doubling the number of novae with measured spectra \cite{Shafter2011}. The fraction of novae belonging to the Fe II and to the He/N class were 82\% and 18\%, respectively, mirroring the population of Galactic novae. The spectral class is related to the speed class, as in alactic novae: He/N novae are faster and more luminous. There is no correlation between the spectral class and the 
position of the nova event in M31.
A spectroscopic survey including eight novae in M33 has been presented by \cite{Shafter2012}. Five novae out of eight belonged to the He/N class, while two of them to the Fe II class, a remarkable difference compared to the Galaxy and M31, dominated by Fe II novae. The large fraction of He/N novae in M33 can be related to its younger stellar population, that leads to more massive white dwarfs than those found in the Galaxy or M31.
The novae in LMC has been investigated by \cite{Shafter2013}, who secured photometric data for 29 novae and spectroscopic data for 18 objects. The fractions of novae belonging to the Fe II and He/N class are very close, in contrast to the dominance of Fe II novae in the Galaxy and in M31, but in broad agreement with the M33 sample. The high fraction of He/N novae can be explained, as in M33, by the young stellar population that leads to massive white dwarfs. Another consequence is the high number of recurrent novae, about 10\%. The He/N novae  in LMC show a faster decline than novae in the Milky Way and in M31.

Some historical novae have been super-luminous, among them the galactic novae V1500 Cyg \cite{DellaValle1991} and the extragalactic LMC 1991 \cite{DellaValle1991}, \cite{Schwarz2001}, SN 2010U in NGC 4214 \cite{Czekala2013}, M31 2007-11d \cite{Shafter2009}, M31 2009-10b \cite{Poggiani2015}. V1500 Cyg probably is an hybrid nova \cite{Shafter2009}, while the extragalactic objects all belong to Fe II class, that should be fainter than He/N novae. The simple correspondence between Fe II novae and older stellar populations with less massive white dwarfs does not necessarily hold; probably the super-luminous novae are systems with cool dwarfs accreting material at a very slow pace \cite{Czekala2013}.

\section{Distances}

The distance of novae are a key parameter to estimate their spatial distribution and their physical parameters. 
The classical methods use the Maximum Magnitude Rate of Decline (MMRD) relations \cite{Cohen1988}, \cite{DellaValleLivio1995}, \cite{DeVaucouleurs1978}, \cite{DownesDuerbeck2000}, \cite{Schmidt1957} or the absolute magnitude achieved 15 days after the maximum \cite{BuscombeDeVaucouleurs1955}, \cite{Capaccioli1989}, \cite{Cohen1985}, \cite{DownesDuerbeck2000}, \cite{VandenberghYounger1987}.

Novae are among the most luminous transients and are good candidates as standard candles to measure extragalactic distances \cite{vandenBerghPritchet1986}, \cite{DellaValleLivio1995}, although novae in distant galaxies are difficult to detect  with medium size telescopes. The Very Large Telescope (VLT) has been used to search for novae in NGC 1316, an early type galaxy in the Fornax cluster \cite{DellaValleGilmozzi2002} at about 20 Mpc. The authors discovered and monitored four novae, estimating the distance of the galaxy with an MMRD relation and the rate of novae in NGC 1316, suggesting that novae can be used as standard indicators and offer an advantage compared to Cepheids, since they can appear in all type of galaxies.

A new class of faint and fast novae has opened a discussion about the validity of the MMRD relations \cite{Kasliwal2011}. The authors have performed a photometric and spectroscopic monitoring of novae  in M 31, M 81, M 82, NGC 2403, NGC 891. The target galaxies were observed every night, producing light curves with high sampling. The parameters of the observed novae were not consistent with the standard MMRD relations, opening a new region of the luminosity vs. decay time diagram. Fast Galactic novae are brighter than slow novae and belong to the He/N spectral class. The observed extragalactic novae belonged to the Fe II spectral class. The nova outburst is governed mainly by the mass of the white dwarf \cite{Livio1992},  but the role of composition, temperature and accretion rate must be included \cite{Kasliwal2011}. The faint and fast novae could contain hot and massive white dwarfs \cite{Kasliwal2011}, as predicted by \cite{Yaron2005}.

The parallax is a primary distance indicator. The first data release of GAIA has produced new estimations for the parallax of 16 cataclysmic variables, among them the three old novae V603 Aql, RR Pic, HR Del \cite{Ramsay2017}.  The GAIA parallaxes of V603 Aql, RR Pic are in broad agreement with the HST parallax estimation. When compared to the measurements based on the shell expansion velocity, the GAIA estimations for RR Pic and HR Del are biased towards slightly smaller distances compared to GAIA. The future releases of GAIA data will provide a large number of parallaxes for other novae and cataclysmic variables in general.

A different estimation method for the distance of Galactic novae has been presented by \cite{Ozdonmez2016}. The approach is based on the location of the red clump giants in the infrared color-magnitude diagram and uses a combination of reddening-distance relations and independent reddening estimates. The method has been calibrated against the distance of novae with shells, with measured expansion parallax \cite{DownesDuerbeck2000}. The authors have estimated the distance of 73 novae and set limits for 46 objects.

\section{Recurrent novae}

To date, there are 10 known recurrent novae in the Galaxy  \cite{Schaefer2010}, 4 in LMC \cite{Orio2015} and 12 in M31 \cite{Shafter2015}. The photometric evolution of all galactic recurrent novae (37 eruption events in total) has been presented by \cite{Schaefer2010}. 

\begin{figure}
\centering
\includegraphics[width=.45\textwidth]{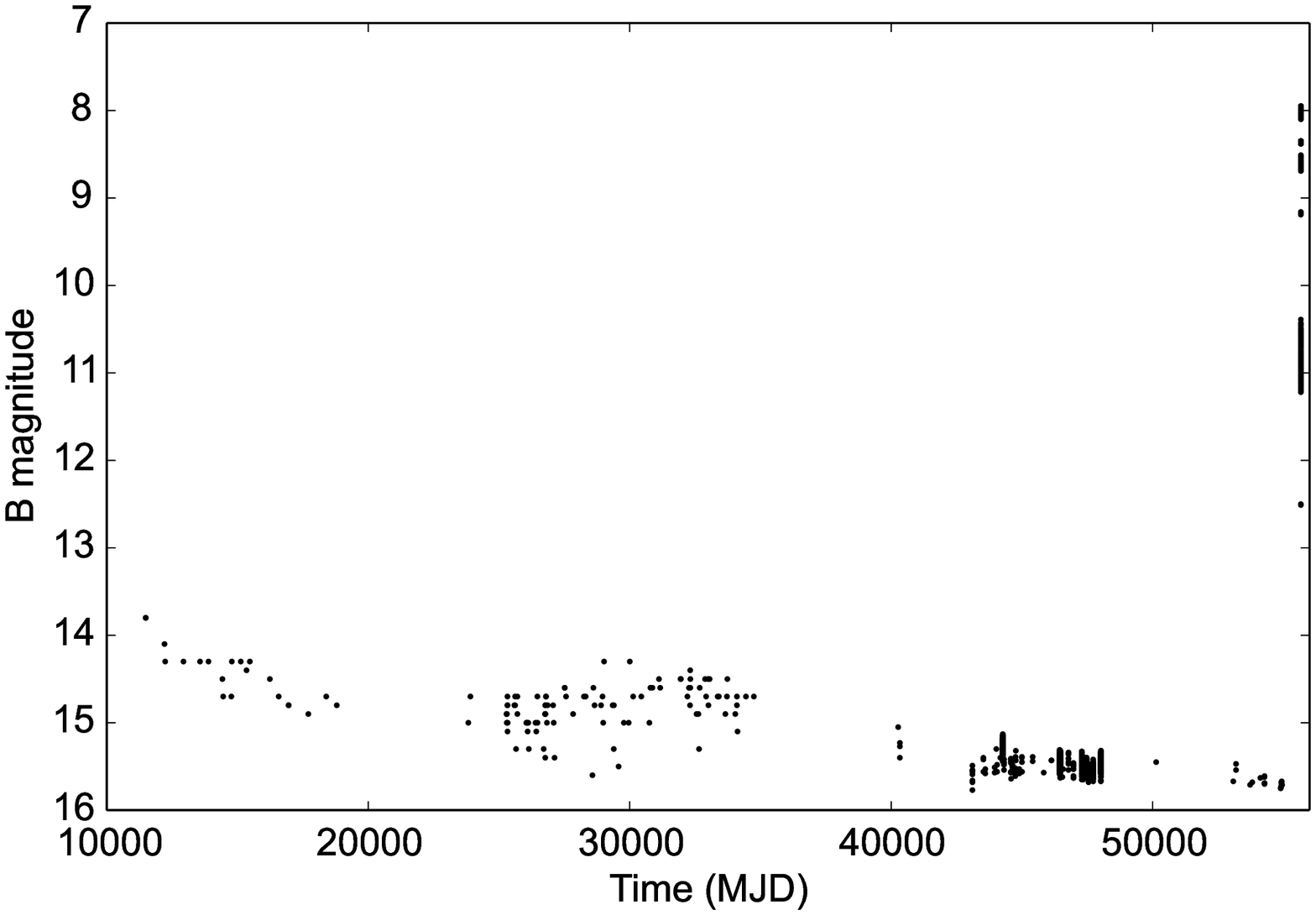}
\includegraphics[width=.45\textwidth]{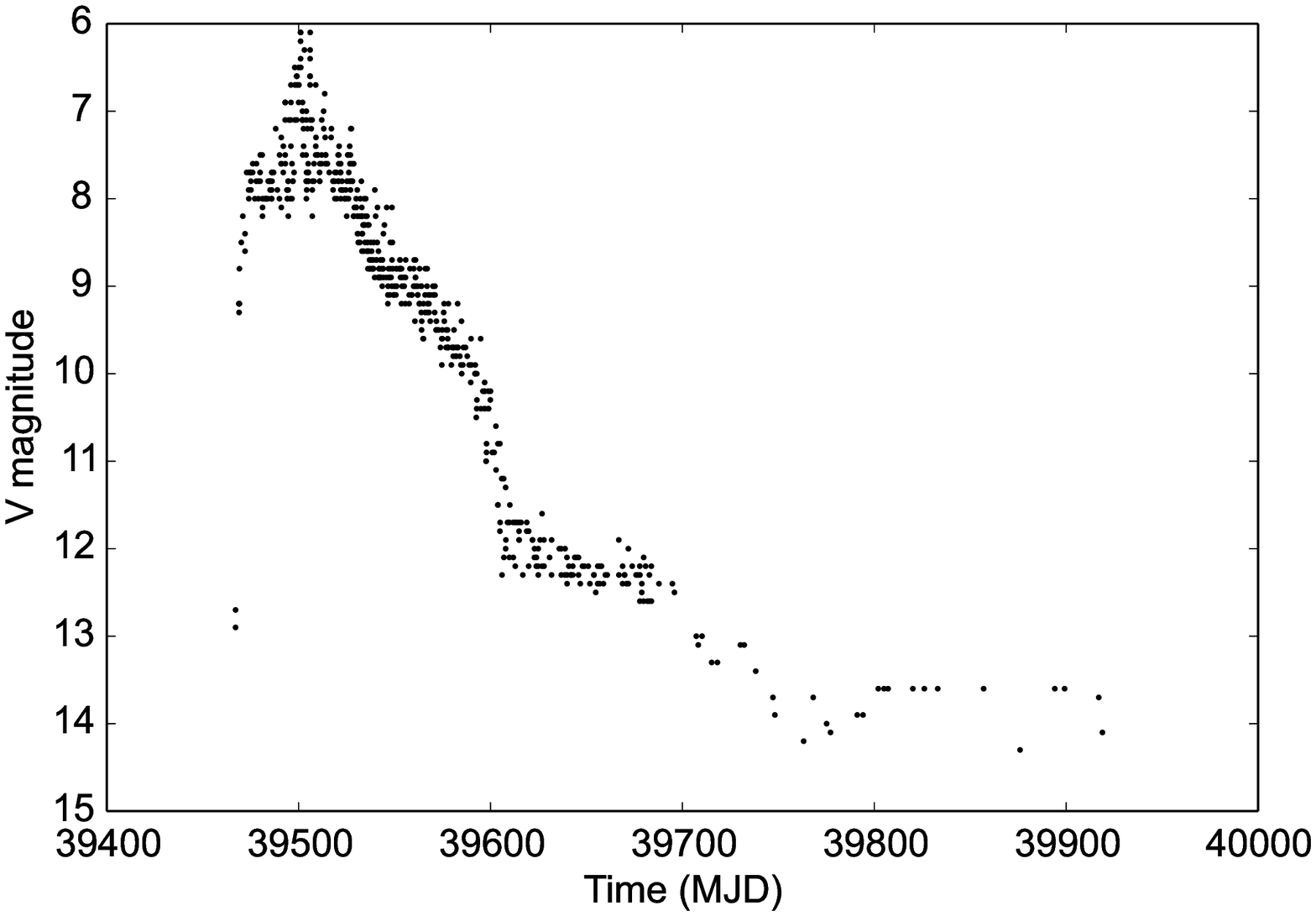}
\caption{The historical light curve (left) and the 1966 outburst (right) of T Pyx; data from \cite{Schaefer2010}}
\label{fig:recurrent}
\end{figure}

An example of the photometric data is shown in Fig. \ref{fig:recurrent} for the historical light curve of T Pyx, that underwent outbursts in 1890, 1902, 1920, 1944, 1966 and 2011 \cite{Schaefer2013}.

Recurrent novae are classified into three classes. The systems T Pyx, IM Nor and CI Aql have an orbital period smaller than one day and secondary stars belonging to main sequence. The systems  U Sco, V394 CrA and V2487 Oph have orbital periods of the order of one day and evolved secondaries. The systems T CrB, RS Oph, V745 Sco and V3890 Sgr have orbital periods of the order of years and red giant secondaries. 
Most recurrent novae show fast decline times and belong to the He/N spectral class.

The fraction of recurrent novae in M31 is about a few percent \cite{Shafter2015}. Among them, nova M31 2008-12a is of special interest, since there are more than ten recorded outbursts, due to the short observed recurrence time, 351$\pm$13 days. The effective period could however be one half of the observed period, 175$\pm$11 days \cite{Henze2015}. The white dwarf of M31 2008-12a  has a mass close to the Chandrasekhar limit, making the nova a strong candidate as a supernova Ia progenitor. The outbursts have triggered multifrequency  observations with ground and space based instruments. The optical, infrared and X-ray monitoring of the 2015 outburst has been reported by \cite{Darnley2016}: the eruptions of 2013-2015 show a similar behavior, suggesting the presence of a red giant donor. The predicted 2016 and 2017 outbursts have been successfully observed \cite{Itagaki2016}, \cite{Boyd2017}.

\section{Infrared studies}

The infrared region of the spectrum provides a great wealth of information about the physical processes in novae. The first observations of FH Ser by \cite{Geisel1970} showed an increase of the flux above a few $\mu$m in correspondence to the fading in the optical light curve, suggesting that the nova had formed dust. The infrared observation of novae can be split between the near infrared and normal infrared regions, at about 3 $\mu$m.

The observations in the near infrared region, involving the J, H, K bands (about 1 to a few microns), have been used to devise a classification system \cite{BanerjeeAshok} in analogy to the system based on the optical post-outburst spectra discussed above. The Fe II and He/N classes appear as separate classes also in the near infrared. Carbon lines appear only in Fe II novae, but not in He/N novae, while hydrogen Paschen and Brackett lines appear in both classes. The richness in carbon lines could be explained by the different hardness level of the radiation by the central remnant after the eruption, the higher content of UV photons could reduce the amount of the neutral carbon producing the above lines \cite{BanerjeeAshok}.
A relevant process in the near infrared is the CO emission, since it could allow the determination of the yield of $^{13}$C and the test of nucleosynthesis models and their impact on the chemical evolution of Galaxy \cite{BanerjeeAshok}. The CO fundamental band lies in the region of the M band, where observations are affected by a strong sky background. The CO emission is studied in the near infrared via the first overtones, starting at 2.29 $\mu$m (K band) \cite{BanerjeeAshok}.

\begin{figure}
\centering
\includegraphics[width=.45\textwidth]{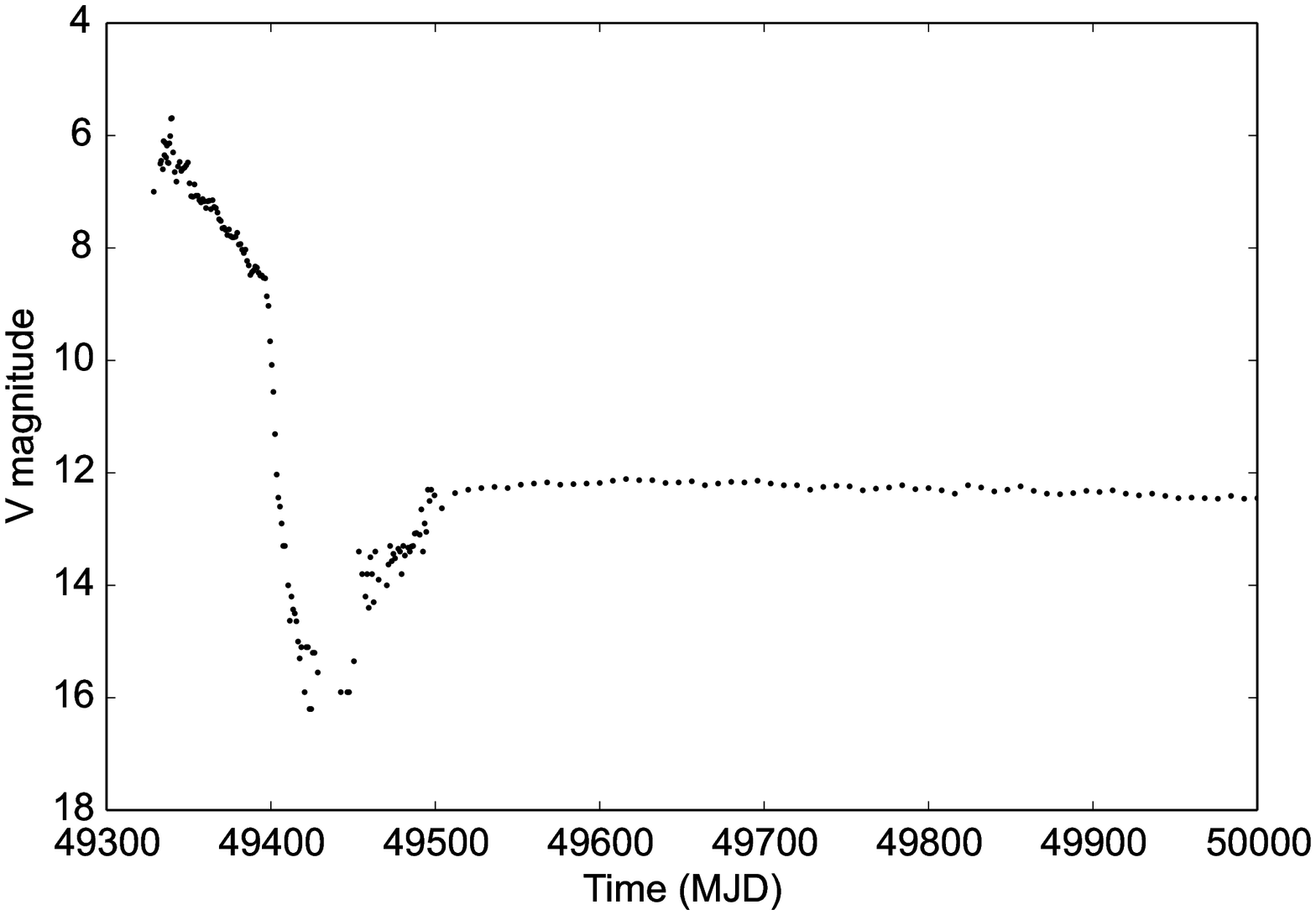}
\includegraphics[width=.45\textwidth]{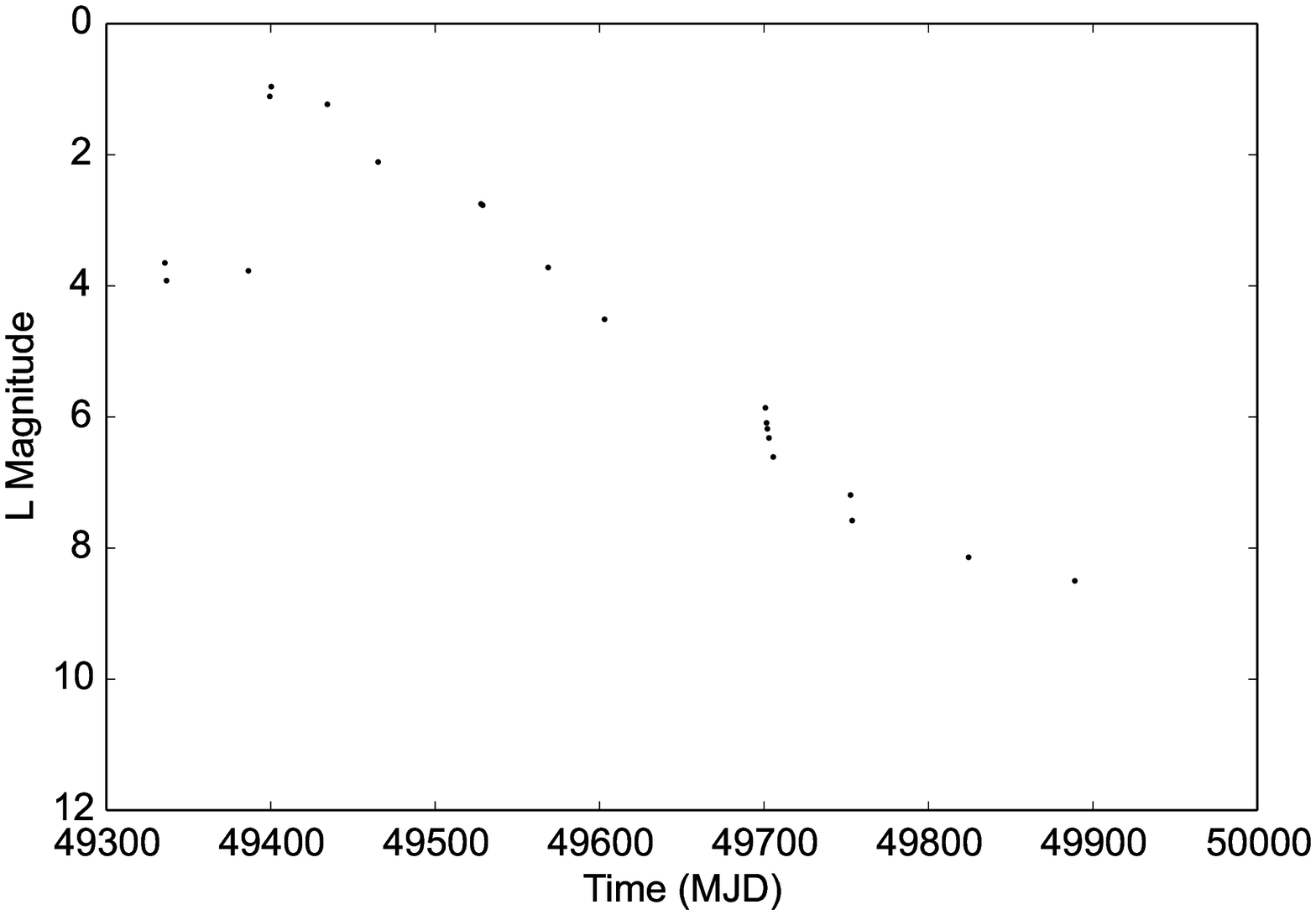}
\caption{The V band light curve (left) and the L band light curve (right) of the dusty nova V705 Cas; optical data from \cite{Strope2010}, infrared data from \cite{Mason1998}}
\label{fig:v705cas}
\end{figure}

The infrared region above 3 $\mu$m \cite{EvansGehrz} is particularly suitable for dust investigations. The composition of dust is variegate, including silicates, amorphous carbon, hydrocarbons, SiC \cite{EvansGehrz}.
The formation of dust is often marked by the appearance of a dip in the optical light curve due to the obscuring effects of dust. The optical fading is accompanied by a strong increase of the infrared radiation, produced by the heating and the re-emission of the dust particles. An example is given by the evolution of V705 Cas, a dust forming nova,  reported in Fig. \ref{fig:v705cas}. The left part of figure presents the optical curve, with an evident dip explained by the formation of an optically thick carbon dust shell \cite{Mason1998}. The infrared (3.6 and 3.8 $\mu$m) light curve shown in the right part achieves the maximum during the dust obscuration undergone by the optical radiation.

\section{Radio studies}

\begin{figure}
\centering
\includegraphics[width=.6\textwidth]{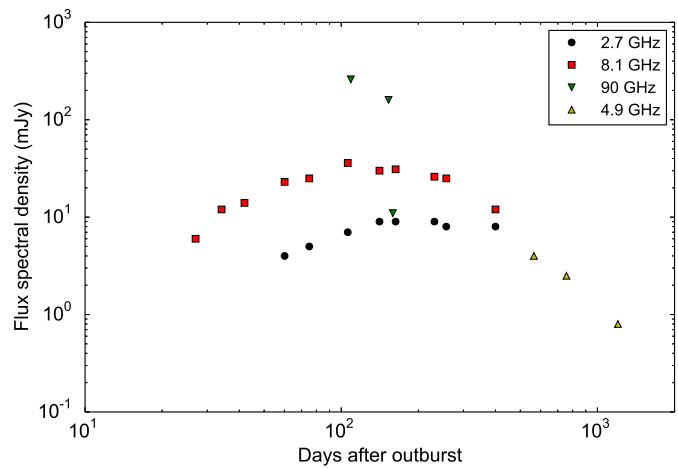}
\caption{The light curves of V1500 Cyg at different frequencies; data from \cite{Hjellming1979}}
\label{fig:v1500cyg}
\end{figure}

The radio emission of novae spans a time interval of years, more extended than that related to the optical emission, of the order of months \cite{Roy2012}, \cite{Kantharia2012}. The main mechanism of radio emission is the thermal bremsstrahlung of the ejecta, that in principle is a tracer of the ejected mass. The radio emission can be explained by a spherical isothermal shell of ionized gas with a density gradient behaving as a power law \cite{BodeEvans2008}. The ejection can be instantaneous, as in the Hubble flow model, or continuous, as in the variable wind model \cite{Roy2012}. The Hubble flow model predicts a quadratic rise during the optically thick stage and an inverse cubic fading during the later optically thin stages. The multifrequency radio curves of V1500 Cyg \cite{Hjellming1979} are reported in Fig. \ref{fig:v1500cyg}.

The E-nova Project \footnote{https://sites.google.com/site/enovacollab/home}  at the Very Large Array (VLA) monitors novae over several frequencies in the region between a few GHz and some hundreds GHz, starting observations within a few days after the eruption. The radio observations are accompanied by optical and X-ray observations. V407 Cyg, the first nova detected in gamma rays, is among the novae included in the program \cite{Chomiuk2012}.

\section{X-ray studies}

Novae are a natural X-ray emitters, since they contain accreting white dwarfs. The first high energy observations date back to EXOSAT \cite{Ogelman1987} and ROSAT \cite{Krautter1996}, which showed intensity variations around nova outbursts and the progressive decrease of the spectrum hardness towards a super-soft phase. The hard and soft components are related to shocks within the ejected material and to the nuclear burning on the surface of the white dwarf. The use of high spectral resolution gratings has allowed the discovery of a large number of emission lines with the observations of XMM-Newton and Chandra \cite{Orio2012}, \cite{Ness2012}. High resolution X-ray spectroscopy is the main tool to obtain information about the temperature and the chemical composition of the burning white dwarf \cite{Orio2012}. The advent of Swift has allowed the production of light curves with high cadence, starting within the first days after the eruption, for several tens novae \cite{Osborne2015}. The Swift observations include novae in the Galaxy, M31 and the Magellanic Clouds.  The combination of Swift and the spectral capabilities of XMM-Newton and Chandra is providing information over different aspects of the nova process.

\begin{figure}
\centering
\includegraphics[width=.6\textwidth]{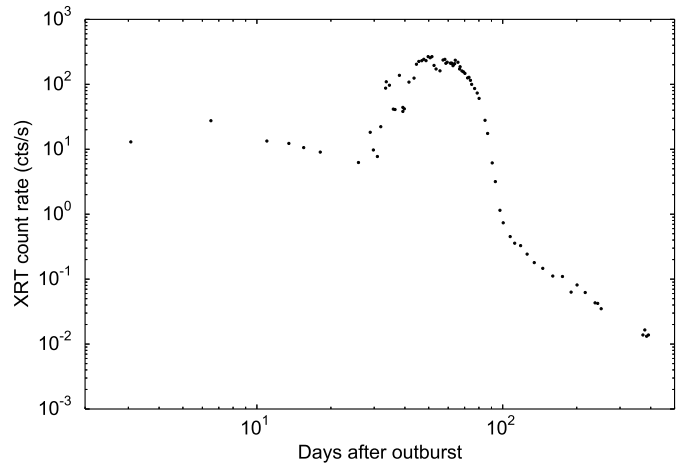}
\caption{The Swift light curve of RS Oph in the 0.3-10 keV band; data from \cite{Schwarz2011}}
\label{fig:rsophx}
\end{figure}

An example of X-ray investigation of a nova is the study of
RS Oph during the outburst that occurred in 2006 and triggered multifrequency observations. The X-ray observations showed the evidence of shocks of ejecta \cite{Bode2006}, \cite{Sokoloski2006}. The shock had a bipolar geometry with an equatorial ring, as shown by the radio interferometric \cite{OBrien2006} and HST observations \cite{Bode2007}. After the outburst, the X-ray flux in the region of a few keV initially declined, before the onset of a cusp around day 30, that marked the beginning of the super-soft phase (Fig. \ref{fig:rsophx}). The drop in the wind density produces a shrinking of the photosphere and the decrease of the optical depth, exposing the hotter layers. The residual burning ended at day 80, when the cusp dropped in intensity.

The spectra before and after the super-soft phase can be modeled with collisional plasmas \cite{Ness2012}. The emission before the onset of the super-soft stage can be explained by shocks, while the emission after the stage is produced by radiatively cooling thin ejected material \cite{Ness2012}. The universal decline law by \cite{HachisuKato2006}, \cite{KatoHachisu1994}  predicts the on and off times of the super-soft phase \cite{HachisuKato2010}.

The sample of novae detected by Swift contains several objects that showed the super-soft phase \cite{Schwarz2011}, allowing a systematic investigation of several processes. The early turn-on and turn-off of the super-soft phase are related to an high ejection velocity, that is related to a fast decline of the optical light curve. The relations holds also for novae in M31 \cite{Henze2014}. The super-soft phase is often correlated with the presence of the forbidden [Fe X] line at 6374~$\AA$ in the optical spectra \cite{Schwarz2011}.
After the end of the super-soft phase, the X-ray flux shows a large variability, including modulations related to the orbital variability or to the rotation period of the white dwarf, and flaring or dimming events \cite{Orio2012}. 

The high resolution X-ray spectra show a variety of behaviours during the super-soft stage, since they can be dominated by the continuum and the absorption features of the white dwarf  or by broad and intense emission lines or by a combination of the previous features \cite{Orio2012}. The previously observed thermal bremsstrahlung continuum started showing broad emission lines with the improvement in the energy resolution of instrumentation \cite{Orio1995}. The main transitions are H-like or He-like. Absorption spectra can suggest the composition of the white dwarf.

\section{Gamma ray studies}

Gamma ray emission in novae had been predicted, but it was expected in the MeV region in association to positron annihilation and nuclear de-excitation in the decays of nitrogen, oxygen and sodium \cite{ClaytonHoyle1974}. A discussion of the corresponding gamma ray emission has been presented by \cite{Hernanz2014}. The first detection of gamma rays from a nova occurred, surprisingly, in the high energy region, when the Fermi-LAT instrument \cite{Atwood2009} observed the radiation of the symbiotic nova V407 Cyg in the region 0.1 to 10 GeV \cite{Abdo2010}.

\begin{figure}
\centering
\includegraphics[width=.7\textwidth]{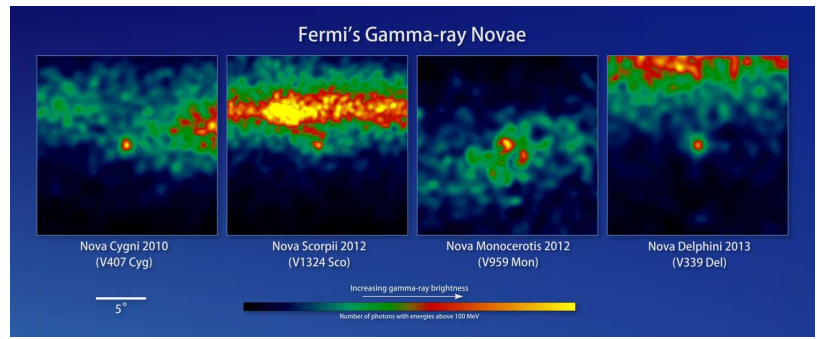}
\caption{Count maps of four new gamma-ray novae observed by Fermi-LAT; adapted from https://fermi.gsfc.nasa.gov/science/eteu/binaries/}
\label{fig:gamma}
\end{figure}

The unexpected GeV emission was caused by the interaction of material in the nova shell with the environmental medium of the red giant secondary. V407 Cyg is a symbiotic nova where ejecta expand inside a dense circumstellar wind, thus particles can be accelerated in a blast wave. After V407 Cyg, the Fermi-LAT has detected GeV gamma ray emission in the classical novae V959 Mon, V1324 Sco and V339 Del \cite{Ackermann2014}. The environment around classical novae has a smaller density than that around symbiotics, thus the acceleration is produced by a bow shock driven by the expelled material in the ISM or many small shocks produced by the presence of turbulence \cite{Ackermann2014}. V959 Mon, V1324 Sco and V339 Del all showed a soft spectrum for a interval of a few weeks. The observation of gamma ray emission in different type of systems suggests that all novae are potential gamma ray emitters. The physical processes involved depend on general properties of the systems, such as the mass of the white dwarf or the mass transfer rate. The main mechanisms for gamma ray emission in novae are the hadronic and the leptonic scenarios \cite{Ackermann2014}. In the hadronic scenario, nova ejecta interact with nuclei in the environment medium (novae) or with the stellar wind (symbiotics), producing neutral pions that decay to photons. In the leptonic scenario, gamma rays are produced by the interaction of accelerated electrons with photons via inverse Compton scattering or with atoms via bremsstrahlung. The Fermi-LAT instrument has observed GeV emission from two additional novae, V1369 Cen and V5668 Sgr \cite{Cheung2016}, detecting gamma rays from two days to tens days after the outburst.  The gamma ray emission showed a longer duration, but was fainter, compared to V959 Mon, V1324 Sco and V339 Del \cite{Ackermann2014}. V1369 Cen and V5668 Sgr showed peculiar outbursts in the optical decline curve during the gamma ray active period. The counting maps of four gamma ray novae observed by Fermi LAT are reported in Fig. \ref{fig:gamma}.

Excluding the symbiotic nova V407 Cyg, the Fermi-LAT instrument has detected gamma rays from six novae, out of the 69 novae detected in the optical domain over the same interval. The discrimination between the hypotheses of gamma ray emission not being a general properties of novae or the faintness of gamma ray emission has been addressed by \cite{Morris2017}, who showed that novae with a magnitude R$\leq$12 and within 8 kpc are potentially detectable by Fermi-LAT \cite{Morris2017}.

Gamma ray novae offer the possibility to investigate the physical mechanisms related to relativistic particle acceleration at non relativistic shocks \cite{Metzger2015}. The high density of ejected material at the epoch of emission of gamma rays produces radiative shocks, the X-rays following the shock are absorbed in neutral gas and processed into optical radiation. 
The fraction of the shock power that is involved in the acceleration of relativistic particles is governed by the relative ratio of gamma ray and optical luminosity, For two novae, V1324 Sco and V339 Del, the ratio is above the value estimated from galactic cosmic rays, favouring hadronic models. Emission of X-ray radiation in the range 10 to 100 keV is predicted to occur in coincidence with GeV emission. 

The investigation of the shocks and the relativistic particle acceleration occurred in the nova outburst \cite{Metzger2016} can involve possible gamma ray emission in the TeV region. The gas upstream the shock is shielded by ionization and is neutral. The acceleration region is confined to a a thin photo-ionized layer ahead the shock. The maximum energy that can be achieved by the above acceleration mechanism is in the range from 10 GeV to 100 TeV. The highest energy gamma rays can be detected by using ground based atmospheric Cherenkov detectors, such as MAGIC or the Cherenkov Telescope Array CTA \cite{Actis2011}. The TeV gamma rays could be associated with TeV neutrinos, that are suitable sources for IceCube. The theoretical modeling of gamma and neutrino emission in gamma ray novae has been presented by \cite{SitarekBednarek2012}.

The ground based atmospheric Cherenkov telescope MAGIC has monitored several novae and dwarf novae in the energy region above 50 GeV \cite{Ahnen2015}. If the GeV emission is produced by the inverse Compton scattering of electrons accelerated in a shock, protons in the same environment could also be accelerated at high energies and contribute to the TeV spectrum. No significant TeV emission was observed in nova V339 Del after its outburst. The interpretation of MAGIC and Fermi data of V339 Del showed that the total power related to accelerated protons was smaller than the 15\% of the total power related to accelerated electrons.

\section{Gravitational wave emission}

The first direct detection of gravitational waves from the binary black hole merger GW150914 \cite{Abbott2016} has opened a new observational window in astronomy, that now includes other black hole mergers and the first coalescence of a binary neutron star system \cite{Abbott2017}.

Cataclysmic variable, including novae, are candidate emitters of gravitational wave in the sub-Hz region, together with massive black hole binaries and stellar mass black holes inspiraling onto supermassive black holes \cite{Mironovskii1966}, \cite{Webbink1987}. The frequency region below one Hz cannot be explored using ground based interferometers, because of seismic noise, but using laser interferometers based in space \cite{Vitale2014}. Space based operation allows to attain much longer arm lengths, of the order of millions km, compared to ground based interferometers. The concept of the space based interferometer LISA has been initially developed by ESA and NASA \cite{LISA}. A gravitational wave produces a tidal acceleration on test masses pairs, measured using the modulation of the frequency of the laser. The initial LISA project proposed a 5 Million km arm length with a system of three spacecrafts containing the free falling test masses and arranged as an equilateral triangle  lagging the Earth by 20 degrees. After the stopping of the NASA support in 2011, LISA was reshaped as eLISA, with a different orbit and a shorter arm length \cite{eLISA}. The new LISA project has been approved in June 2017 and is scheduled for launch after 2030. The three spacecrafts of LISA (mother and two daughters) will orbit around the Sun in nearly circular orbits, forming an interferometer with one to two million km arm length. The spacecrafts contain test masses in free fall that are shielded from external disturbances. The spacecraft is centered on the test mass with a group of thrusters. The acceleration of the spacecraft relative to the test mass is measured and subtracted.

The LISA interferometer has been preceded by the LISA Pathfinder, the demonstrator of the future instrument and of the related technologies \cite{pathfinder1}. The Pathfinder contains a small version of the  final LISA arm, with two test masses (2 kg Au-Pt cubes) at a distance of 0.38 m. The LISA Pathfinder has been launched in December 2015, successfully demonstrating the feasibility of free-falling test masses in orbit and showing a noise smaller than a factor 5 than the initial design \cite{pathfinder2}, close to the specifications of the original LISA design.

Different noises contribute to the sensitivity of space based interferometers. The region below milliHz is dominated by the acceleration noise, the product of residual forces acting on the test masses. The  region above some tens mHz is dominated by the shot noise of the laser. Space based interferometers show also a noise produced by a background with astrophysics origin, the confusion noise of  the unresolved population of galactic binaries \cite{Hils1990}, \cite{Bender1997}.

Cataclysmic variables emit gravitational waves at twice the orbital frequency and at the corresponding harmonics, as other binary systems. Since orbits become more and more circularized in time, the contribution of the harmonics is negligible. The gravitational wave strain produced by a binary system is given by \cite{Thorne1987}:

\begin{equation}
h = 8.7 \times 10^{-21} \left( \frac{\mu}{M_{\odot}} \right ) \left ( \frac{M}{M_{\odot}} \right )^{\frac{2}{3}}
\left ( \frac{100 pc}{r} \right ) \left ( \frac{f}{10^{-3} Hz} \right ) ^{\frac{2}{3}}
\end{equation}

where $M=M_1+M_2$, $\mu = \frac{M_1 M_2}{M_1+M_2}$,  $M_1$, $M_2$ are the masses of the primary and secondary star, $r$ the cataclysmic distance and $f$ the frequency. 

\begin{figure}
\centering
\includegraphics[width=.7\textwidth]{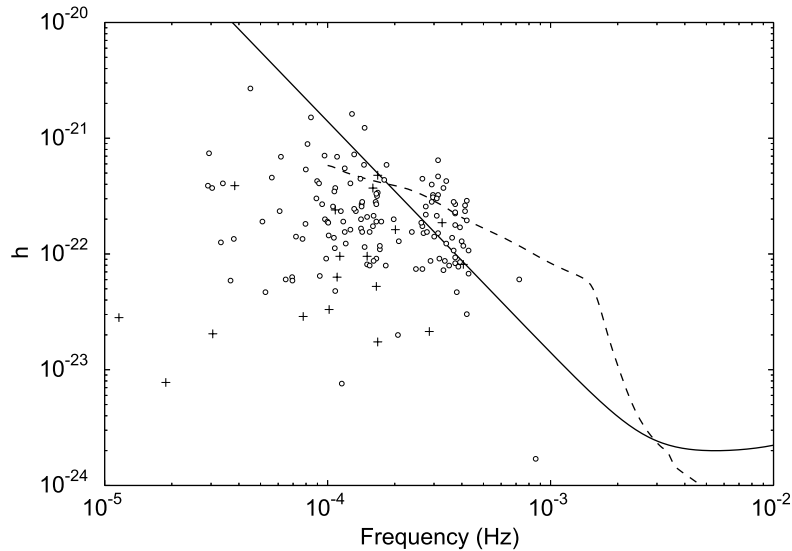}
\caption{Gravitational wave emission of novae (crosses) and of non novae cataclysmic variables (empty circles), data from \cite{Meliani2000}; the solid curve is the instrumental sensitivity of the LISA interferometer, while the dashed line is the binary confusion noise, where both noise curves have been estimated assuming 1 year of integration time and unit signal to noise ratio \cite{Amaro2012}}
\label{fig:gw}
\end{figure}

The strain $h$ is reported in Fig. \ref{fig:gw} for the sample of 156 cataclysmic variables investigated by \cite{Meliani2000}, split into novae (crosses) and non novae cataclysmic variables (empty circles). The solid curve is the sensitivity of the LISA interferometer \cite{Amaro2012}. The dashed curve is the contribution of the confusion noise \footnote{http://www.srl.caltech.edu/$~$shane/sensitivity/}, the astrophysical background produced by unresolved binary systems \cite{Hils1990}, \cite{Bender1997}. The instrumental noise and the confusion noise have been estimated for 1 year of integration time and unit signal to noise ratio.

Novae are not the easily detectable sources, compared to other cataclysmic variables, because of their long orbital periods.

\section{Conclusions}

Classical and recurrent novae are  multifrequency laboratories that allow the exploration of a rich variety of processes involved in their eruption. The historical photometric and spectroscopic observations in the optical domain are now performed with high temporal cadence. The evolution of X-ray and gamma ray instrumentation has allowed the investigation of new phenomena, such as the rich patterns of X-ray spectra and the unexpected GeV emission. The high luminosity of novae makes them potential standard candles, since they can be observed up to a distance of tens Mpc.

\acknowledgments

The author thanks the organizers for the invitation to the workshop. Many thanks to the Telescope Allocation Time of the Loiano Observatory for the observing time. The author is grateful to the referee for the careful comments.


\begin{thebibliography}{999}
\bibitem{Abbott2016} B. P. Abbott et al., \emph{PRL} {\bf 116} (2016) 061102.
\bibitem{Abbott2017} B. P. Abbott et al., \emph{PRL} {\bf 119} (2017) 161101.
\bibitem{Abdo2010} A. A. Abdo et al., \emph{Sci} {\bf 329} (2010) 817.
\bibitem{Ackermann2014} M. Ackermann et al., \emph{Sci} {\bf 345} (2014) 554.
\bibitem{Actis2011} M. Actis et al., \emph{ExA} {\bf 32} (2001) 143.
\bibitem{Ahnen2015} M. L. Ahnen et al., \emph{A\&A} {\bf 582} (2015) 582.
\bibitem{Allen1954} C. W. Allen, \emph{MNRAS} {\bf 114} (1954) 387.
\bibitem{Amaro2012} P. Amaro-Seoane et al., \emph{CQG} \textbf{29} (2012) 124016.
\bibitem{eLISA} P. Amaro-Seoane et al., \emph{GW Notes} \textbf{6} (2013) 4.
\bibitem{pathfinder1} F. Antonucci et al., \emph{CQG} \textbf{29} (2012) 124014.
\bibitem{pathfinder2} M. Armano et al., \emph{PRL} {\bf 116} (2016) 231101.
\bibitem{Arp1956} H. C. Arp et al., \emph{AJ} {\bf 61} (1956) 15.
\bibitem{Atwood2009} W. B. Atwood et al., \emph{ApJ} {\bf 697} (2009) 1071.
\bibitem{BanerjeeAshok} D. P. K. Banerjee and N. M. Ashok, \emph{BASI} {\bf 40} (2012) 243.
\bibitem{Bender1997} P. L. Bender and D. Hils, \emph{CQG} \textbf{14} (1997) 1439.
\bibitem{Bode2006} M. F. Bode et al., {\emph ApJ} {\bf 652} (2006) 629.
\bibitem{Bode2007} M. F. Bode at al., {\emph ApJ} {\bf 665} (2007) L63.
\bibitem{BodeEvans2008} M. F. Bode and A. Evans, \emph{Classical Novae}, Cambridge University Press (2012).
\bibitem{Boyd2017} D. Boyd et al., \emph{ATel} {\bf 11116} (2017).
\bibitem{BuscombeDeVaucouleurs1955} W. Buscombe and G. de Vaucouleurs, \emph{Obs} {\bf  75} (1955) 170.
\bibitem{Cao2012} Y. Cao et al., \emph{ApJ} {\bf 752} (2012) 133.
\bibitem{Capaccioli1989} M. Capaccioli et al., \emph{AJ} {\bf 97} (1989) 1622.
\bibitem{Capaccioli1990} M. Capaccioli et al., \emph{ApJ} {\bf 360} (1990) 63.
\bibitem{CassatellaViotti1990} A. Cassatella and R. Viotti, \emph{Physics of Classical Novae}, Springer-Verlag, Lecture Notes in Physics 369 (1990).
\bibitem{Cheung2016} C. C. Cheung et al., \emph{ApJ} {\bf 826} (2016) 142.
\bibitem{Chomiuk2012} L. Chomiuk et al., \emph{ApJ} {\bf 761} (2012) 173.
\bibitem{Ciardullo1987} R. Ciardullo et al., \emph{ApJ} {\bf 318} (1987) 520.
\bibitem{Ciardullo1990} R. Ciardullo et al., \emph{ApJ} {\bf 356} (1990) 472.
\bibitem{ClaytonHoyle1974} D. D. Clayton and F. Hoyle, \emph{ApJ} {\bf 187} (1974) 101.
\bibitem{Cohen1985} J. G. Cohen, \emph{ApJ} {\bf 292} (1985) 90.
\bibitem{Cohen1988} J. G. Cohen, in: \emph{The extragalactic distance scale}, \emph{ASPC} {\bf 4} (1988) 114.
\bibitem{Collazzi2009} A. C. Collazzi et al., \emph{AJ} {\bf 138} (2009) 1846.
\bibitem{Czekala2013} I. Czekala et al., \emph{ApJ} {\bf 765} (2013) 57.
\bibitem{Darnley2006} M. J. Darnley et al., \emph{MNRAS} {\bf 369} (2006) 257.
\bibitem{Darnley2016} M. J. Darnley et al., \emph{ApJ} {\bf 833} (2016) 149.
\bibitem{DeVaucouleurs1978} G. de Vaucouleurs, \emph{ApJ} {\bf 223} (1978) 351.
\bibitem{DellaValle1988} M. Della Valle, in: \emph{The extragalactic distance scale}, \emph{ASPC} {\bf 4} (1988) 73.
\bibitem{DellaValle1992} M. Della Valle et el., \emph{A\&A} {\bf 266} (1992) 232. 
\bibitem{DellaValle1994} M. Della Valle et al., \emph{A\&A} {\bf 287} (1994) 403.
\bibitem{DellaValle1991} M. Della Valle, \emph{A\&A} {\bf 252} (1991) L9.
\bibitem{DellaValle2002} M. Della Valle, in \emph{Classical Nova Explosions}, \emph{AIPC} {\bf 637} (2002) 443.
\bibitem{DellaValleGilmozzi2002} M. Della Valle and R. Gilmozzi, \emph{Sci} {\bf 296} (2002) 1275.
\bibitem{DellaValleLivio1994} M. della Valle and M. Livio, \emph{A\&A} {\bf 286} (1994) 786.
\bibitem{DellaValleLivio1995} M. della Valle and M. Livio, \emph{ApJ} {\bf 452} (1995) 704.
\bibitem{DellaValleLivio1998} M. della Valle and M. Livio, \emph{ApJ} {\bf 506} (1998) 818.
\bibitem{DownesDuerbeck2000} R. A. Downes and H. W. Duerbeck, \emph{AJ} {\bf 120} (2000), 2007.
\bibitem{Downes2005} R. A. Downes et al., \emph{JAD} {\bf 11} (2005) 2
\bibitem{Duerbeck1987} H. W. Duerbeck, \emph{SSR} {\bf 45} (1987) 1. \bibitem{Duerbeck1992} H. W. Duerbeck, \emph{MNRAS} {\bf 258} (1992) 629.
\bibitem{Duerbeck2009} H. W. Duerbeck, \emph{AN} {\bf 330} (2009) 568.
\bibitem{Ederoclite2014} A. Ederoclite, in: \emph{Stella Novae: Past and Future Decades},  \emph{ASPC} {\bf 490} (2014) 163.
\bibitem{EvansGehrz} A. Evans and R. D. Gehrz, \emph{BASI} {\bf 40} (2012) 213.
\bibitem{Geisel1970} S. L. Geisel et al., {\emph ApJ} {\bf 161} (1970) L101. 
\bibitem{Giovannelli2015} F. Giovannelli and L. Sabau-Graziati, in
Proceedings of \emph{The Golden Age of Cataclysmic Variables and Related Objects - III}, \pos{PoS(Golden2015)001} (2016).
\bibitem{Graham1979} J. A. Graham, in \emph{Changing Trends in Variable Star Research}, \emph{IAU Coll.} {\bf 46} (1979) 96.
\bibitem{HachisuKato2006} I. Hachisu and M. Kato, \emph{ApJSS} {\bf 167} (2006) 59.
\bibitem{HachisuKato2010} I. Hachisu and M. Kato, \emph{ApJ} {\bf 709} (2010) 680.
\bibitem{Hatano1997} K. Hatano et al., \emph{MNRAS} {\bf 290} (1997) 113.
\bibitem{Henze2014} M. Henze et al., {\emph A\&A} {\bf 563} (2014) A2.
\bibitem{Henze2015} N. Henze et al., \emph{A\&A} {\bf 582} (2015) 8.
\bibitem{Hernanz2014} M. Hernanz, in \emph{Stella Novae: Past and Future Decades}, \emph{ASPC} {\bf 490} (2014) 319.
\bibitem{Hils1990} D. Hils et al., \emph{ApJ} \textbf{360} (1990) 65.
\bibitem{Hjellming1979} R. M.Hjellming et al., \emph{AJ} {\bf 84} (1979) 1619.
\bibitem{Hounsell2010} R. Hounsell et al., \emph{ApJ} {\bf 734} (2010) 480.
\bibitem{Hounsell2016} R. Hounsell et al., \emph{ApJ} {\bf 820} (2016) 104.
\bibitem{Hubble1929} E. P. Hubble, \emph{ApJ} {\bf 69} (1929) 103.
\bibitem{Hudec2015} R. Hudec, in Proceedings of \emph{The Golden Age of Cataclysmic Variables and Related Objects - III}, \pos{PoS(Golden2015)041} (2016).
\bibitem{ImamuraTanabe2012} K. Imamura and K. Tanabe, \emph{PASJ} {\bf 64} (2012) L9.
\bibitem{Itagaki2016} K. Itagaki et al., \emph{ATel} {\bf 9848} (2016).
\bibitem{Izzo2015} L. Izzo et al., \emph{ApJL} {\bf 808} (2015) L14.
\bibitem{Kantharia2012} N. G. Kantharia, \emph{BASI} {\bf 40} (2012) 311.
\bibitem{Kasliwal2011} M. M. Kasliwal et al., \emph{ApJ} {\bf 735} (2011) 94.
\bibitem{KatoHachisu1994} M. Kato and I. Hachisu, \emph{ApJ} {\bf 437} (1994) 803.
\bibitem{Krautter1996} J. Krautter et al., \emph{ApJ} {\bf 456} (1996) 788.
\bibitem{LillerMayer1987} W. Liller and B. Mayer, \emph{PASP} {\bf 99} (1987) 600.
\bibitem{LISA} LISA Study Team, 1998, in \emph{LISA Pre-Phase A Report. 2nd Edition}, \textbf{Publication MPQ-233} (1998) Max-Plank Institute for Quantum Optics, Garching.
\bibitem{Livio1992} M. Livio, \emph{ApJ} {\bf 393} (1992) 516.
\bibitem{Mason1998} C.G. Mason et al., {\emph ApJ} {bf 494} (1998) 783.
\bibitem{Matteucci2003} F. Matteucci et al., \emph{A\&A} {\bf 405} (2003) 23.
\bibitem{Meliani2000} M. T. Meliani et al., \emph{A\&A} \textbf{358} (2000) 417.
\bibitem{Metzger2015} B. D. Metzger et al., \emph{MNRAS} {\bf 450} (2015) 2739.
\bibitem{Metzger2016} B. D. Metzger et al., \emph{MNRAS} {\bf 457} (2016) 1786.
\bibitem{Mironovskii1966} V. N. Mironovskii, \emph{SvA} {\bf 9} (1966) 752.
\bibitem{Morris2017} P. J. Morris et al., \emph{MNRAS} {\bf 495} (2017) 1218.
\bibitem{Mroz2015} P. Mroz et al., \emph{ApJS} {\bf 219} (2015) 26.
\bibitem{Mroz2016a} P. Mroz et al., \emph{ApJS} {\bf 232} (2016a) 9.
\bibitem{Mroz2016b} P. Mroz et al., \emph{Nat} {\bf 537} (2016b) 649.
\bibitem{Mukai} https://asd.gsfc.nasa.gov/Koji.Mukai/novae/novae.html
\bibitem{Ness2012} J. U. Ness, \emph{BASI} {\bf 40} (2012) 353.
\bibitem{OBrien2006} T. J. O'Brien et al., {\emph Nat} {\bf 442} (2006) 279.
\bibitem{Ogelman1987} H. \"Ogelman et al., {\emph A\&A} {\bf 177} (1987) 110.
\bibitem{Orio1995} M. Orio et al., \emph{ApJ} {\bf 466} (1995) 410.
\bibitem{Orio2012} M. Orio, \emph{BASI} {\bf 40} (2012) 333.
\bibitem{Orio2015} M. Orio, in Proceedings of \emph{The Golden Age of Cataclysmic Variables and Related Objects - III}, \pos{PoS(Golden2015)064} (2016).
\bibitem{Osborne2011} J. P. Osborne eta al., \emph{ApJ} {\bf 727} (2011) 124.
\bibitem{Osborne2015} J. P.Osborne, \emph{JHEA} {\bf 7} (2015) 117.
\bibitem{Ozdonmez2016} A. \"Ozd\"onmez et al., \emph{MNRAS} {\bf 461} (2016) 1177.
\bibitem{Pagnotta2017} A. Pagnotta,  in \emph{20th European White Dwarf Workshop}, \emph{ASPC} {\bf 509} (2017) 535.
\bibitem{Patterson2014} J. Patterson, \emph{SASS} {\bf 33} (2014) 17.
\bibitem{PayneGaposchkin1957} C. H. Payne-Gaposchkin, \emph{The Galactic Novae}, North-Holland Pub. Co. (1957).
\bibitem{Pietsch2010} W. Pietsch, \emph{AN} {\bf 331} (2010) 187.
\bibitem{Poggiani2006} R. Poggiani, \emph{AN} {\bf 327} (2006) 895.
\bibitem{Poggiani2008a} R. Poggiani, \emph{NewA} {\bf 13} (2008a) 557.
\bibitem{Poggiani2008b} R. Poggiani, \emph{Ap\&SS} {\bf 315} (2008b) 79.
\bibitem{Poggiani2009a} R. Poggiani, \emph{NewA} {\bf 14} (2009a) 4.
\bibitem{Poggiani2009b} R. Poggiani, \emph{AN} {\bf 330} (2009b) 77.
\bibitem{Poggiani2009c} R. Poggiani, \emph{Ap\&SS} {\bf 323} (2009c) 319.
\bibitem{Poggiani2010a} R. Poggiani, \emph{NewA} {\bf 15} (2010a) 657.
\bibitem{Poggiani2010b} R. Poggiani, \emph{NewA} {\bf 15} (2010b) 170.
\bibitem{Poggiani2011} R. Poggiani, \emph{Ap\&SS} {\bf 333} (2011) 115.
\bibitem{Poggiani2012} R. Poggiani, \emph{MmSAI} {\bf 83} (2012) 753.
\bibitem{Poggiani2015} R. Poggiani, \emph{NewA} {\bf 37} (2015) 9.
\bibitem{Poggiani2015a} R. Poggiani, \emph{AcPPP} {\bf 2} (2015) 234.
\bibitem{Ramsay2017} G. Ramsay et al., \emph{A\&A} {\bf 604} (2017) A107.
\bibitem{RitterKolb2003} H. Ritter and U. Kolb, \emph{A\&A} {\bf 404} (2003) 301.
\bibitem{RitterKolb} http://wwwmpa.mpa-garching.mpg.de/RKcat/
\bibitem{Robinson1975} E. L. Robinson, \emph{AJ} {\bf 80} (1975) 515.
\bibitem{Rosino1964} L. Rosino, \emph{AnAp} {\bf 27} (1964) 498.
\bibitem{Rosino1973} L. Rosino, \emph{A\&AS} {\bf 9} (1973) 347.
\bibitem{Rosino1989} L. Rosino, \emph{AJ} {\bf 97} (1989) 83.
\bibitem{Roy2012} N. Roy et al., \emph{BASI} {\bf 49} (2012) 293.
\bibitem{Schaefer2010} B. E. Schaefer, \emph{ApJS} {\bf 187} (2010) 275.
\bibitem{Schaefer2013} B. E. Schaefer et al., \emph{ApJ} {\bf 773} (2013) 55. 
\bibitem{Schmidt1957} T. Schmidt, \emph{Z. Astrophys.} {\bf 41} (1957) 182.
\bibitem{Schwarz2001} G. J. Schwarz et al., \emph{MNRAS} {\bf 320} (2001) 103.
\bibitem{Schwarz2011} G. Schwarz et al., {\emph ApJSS} {\bf 197} (2011) 31.
\bibitem{Shafter1997} A. W. Shafter, \emph{ApJ} {\bf 487} (1997) 226.
\bibitem{Shafter2002} A. W. Shafter, in: \emph{Classical Nova Explosions: International Conference on Classical Nova Explosions}, \emph{AIPC} {\bf 637} (2002) 462.
\bibitem{Shafter2009} A. W. Shafter et al., \emph{ApJ} {\bf 690} (2009) 1148
\bibitem{Shafter2011} A. W. Shafter et al., \emph{ApJ} {\bf 734} (2011) 12.
\bibitem{Shafter2012} A. W. Shafter et al., \emph{ApJ} {\bf 752} (2012) 156.
\bibitem{Shafter2013} A. W. Shafter, \emph{AJ} {\bf 145} (2013) 117.
\bibitem{Shafter2014} A. W. Shafter, in: \emph{Stella Novae: Past and Future Decades}, \emph{ASPC} {\bf 490} (2014) 77.
\bibitem{Shafter2015} A. W. Shafter et al., \emph{ApJS} {\bf 216} (2015) 34.
\bibitem{Shafter2017} A. W. Shafter, \emph{ApJ} {\bf 834} (2017) 196.
\bibitem{ShafterIrby2001} A. W. Shafter and P. K. Irby, \emph{ApJ} {\bf 563} (2001) 749.
\bibitem{Shara1986} M. M. Shara et al., \emph{ApJ} {\bf 311} (1986) 163.
\bibitem{Shara2007} M. M. Shara et al., \emph{Nat} {\bf 446} (2007) 159.
\bibitem{Shara2012} M.M. Shara et al.,\emph{ApJ} {\bf 758} (2012) 121.
\bibitem{Sharov1972} A. S. Sharov, \emph{SvA} {\bf 16} (1972) 41.
\bibitem{Sharov1993} A. S. Sharov, \emph{AstL} {\bf 19} (1993) 147.
\bibitem{SitarekBednarek2012} J. Sitarek and W. Bednarek, \emph{PRD} {\bf 86} (2012) 063011.
\bibitem{Sokoloski2006} J. J. Sokoloski et al., {\emph Nat} {\bf 442} (2006) 276.
\bibitem{Starrfield2016} S. Starrfield et al., \emph{PASP} {\bf 128} (2016) 051001.
\bibitem{Strope2010} R. J. Strope et al., \emph{AJ} {\bf 140} (2010) 34.
\bibitem{Surina2011} F. Surina et al., arXiv:1111.5524 (2011).
\bibitem{Tanaka2011} J. Tanaka et al., \emph{PASJ} \textbf{63} (2011) 911.
\bibitem{Tappert2012} C. Tappert et al., \emph{MNRAS} {\bf 423} (2012) 2476.
\bibitem{Tappert2015} C. Tappert et al., in: Proceedings of \emph{The Golden Age of Cataclysmic Variables and Related Objects - III}, \pos{PoS(Golden2015)062} (2016).
\bibitem{Thorne1987} K. S. Thorne, in \emph{Three Hundreds Years of Gravitation} (1987) 330, Cambridge University Press, Cambridge, eds. S. Hawking and W. Israel.
\bibitem{TomaneyShafter1992} A. B. Tomaney and A. W. Shafter,  \emph{ApJS} {\bf 81} (1992) 683.
\bibitem{vandenBerghPritchet1986} S. van den Bergh and C. J. Pritchet, \emph{PASP} {\bf 98} (1986) 110.
\bibitem{VandenberghYounger1987} S. van den Bergh and P. F. Younger, \emph{A\&AS} {\bf 70} (1987) 125.
\bibitem{vandenBergh1991} S. van den Bergh, \emph{PASP} {\bf 103} (1991) 609.
\bibitem{Vitale2014} S. Vitale, \emph{GRG} \textbf{46} (2014) 1730. 
\bibitem{Walter2012} F. M. Walter et al., \emph{PASP} {\bf 124} (2012) 1057.
\bibitem{Walter2014} F. M. Walter, in \emph{Stella Novae: Past and Future Decades}, \emph{ASPC} {\bf 490} (2014) 191.
\bibitem{Webbink1987} R. F. Webbinbk et al., \emph{ApJ} {\bf 314} (1987) 653.
\bibitem{Williams1991} R. E. Williams et al., \emph{ApJ} {\bf 376} (1991) 721.
\bibitem{Williams1992} R. E. Williams, \emph{AJ} {\bf 104} (1992) 725.
\bibitem{Williams1994} R. E. Williams et al., \emph{ApJSS} {\bf 90} (1994) 297.
\bibitem{Williams2003} R. E. Williams, \emph{JAD} {\bf 9} (2003).
\bibitem{Williams2004} R. E. Williams and A. W. Shafter, \emph{ApJ} {\bf 612} (2004) 867.
\bibitem{Williams2012} R. Williams, \emph{AJ} {\bf 144} (2012) 98.
\bibitem{Yaron2005} O. Yaron et al., \emph{ApJ} {\bf 623} (2005) 398.
\bibitem{Yungelson1997} L. Yungelson et al., \emph{ApJ} {\bf 481} (1997) 127.

\end{thebibliography}
\end{document}